\documentstyle[preprint,aps,epsf]{revtex}

\def\bl{{\bf l}}
\def\p{{\bf p}}
\def\a{{\bf a}}
\def\q{{\bf q}}
\def\k{{\bf k}}
\def\x{{\bf x}}
\def\y{{\bf y}}
\def\A{{\bf A}}
\def\B{{\bf B}}
\def\D{{\bf D}}

\def\h{{1\over 2}}

\begin{document}
\title{
%\hfill \parbox[t]{2 in} {\rm \small hep-ph/9604xxx}
%\vskip 1.3 cm 
 From Current to Constituent Quarks: a Renormalization Group Improved 
Hamiltonian-based Description of Hadrons.}

\author{Adam P. Szczepaniak and Eric S. Swanson}

\address{
   Department of Physics, 
   North Carolina State University,
   Raleigh, North Carolina 27695-8202}

\maketitle

\begin{abstract}
A model which combines the perturbative behavior of QCD with low energy 
phenomenology in a unified framework is developed.
This is achieved by applying a similarity 
transformation to the QCD Hamiltonian which removes interactions between the
ultraviolet cutoff and an arbitrary lower scale. Iteration then yields a 
renormalization group improved effective Hamiltonian at the hadronic energy 
scale.
The procedure preserves the standard ultraviolet behavior of QCD. 
Furthermore, the Hamiltonian evolves smoothly to a phenomenological low
energy behavior below the hadronic scale. This method has the benefit of
allowing radiative corrections to be directly incorporated into nonperturbative
many-body techniques. It is applied to Coulomb gauge QCD supplemented with a 
low energy linear confinement interaction. A nontrivial vacuum is included in
the analysis via a Bogoliubov-Valatin transformation. Finally, the formalism is 
applied to the vacuum gap equation, the quark condensate, and the dynamical 
quark mass. 
\end{abstract}
\date{September, 1996}
\narrowtext

\section{Introduction} 

There is currently an unfortunate dichotomy in analytical approaches to QCD. 
One
either calculates using Lagrangian methods and, with the exception of Monte 
Carlo simulations, is limited to perturbative calculations,  
 or one employs low energy models of
QCD and eschews the known high energy behavior of the theory.  
There are only a few approaches where, for example, 
the calculation of hadronic
form factors may be connected to the expected quark counting behavior at
high energies. These include 
QCD sum rules~\cite{SVZ}, 
NRQCD~\cite{PL}, and to some extent 
light cone quantization~\cite{OSU}. 
This paper develops a formalism by which the known high energy
behavior of QCD (pQCD) may be rigorously joined to standard 
phenomenological models 
of hadrons.   
This is achieved by using the renormalization group to run the scale of the
QCD Hamiltonian to the hadronic regime where it smoothly joins a fixed 
phenomenological behavior.

This approach imposes important constraints on the low
energy portion of the model. If one wishes to recover pQCD, one must work with
current quarks. Thus the low energy theory must also deal with current quarks.
However phenomenology indicates that constituent quarks are the relevant degrees
of freedom at low energy.
Thus, one must allow for spontaneous chiral symmetry breaking in the model.
This, in turn, means that a nontrivial vacuum must be incorporated into the
theory. Finally, we see that one must use many-body techniques when 
calculating observables. 
The Hamiltonian-based renormalization group we employ is particularly useful in
this regard because it allows for the use of nonperturbative many-body 
techniques. 

We call this approach the ``Dynamical Quark Model'' (DQM). The fact that the 
low energy behavior of the DQM evolves directly from pQCD is very helpful 
when analyzing the dynamical structure of the low energy portion of the 
model. For example, we are able to rigorously establish that the non-Abelian
Coulomb interaction is responsible for the quark structure of confinement in the
heavy quark limit. We are also able to resolve ambiguities  
in the separation of short and long distance phenomena when evaluating 
 current matrix elements. 

This brings us to a final important reason for employing the renormalization
group improved (RGI) Hamiltonian. As stated, we are obliged to use many-body 
methods while 
working with the model.
The major tools at our disposal are the Tamm-Dancoff approximation (TDA) and 
the random phase approximation (RPA). These build, for example, meson bound 
states of
the type $q \bar q$ or $q \bar q + q \bar q (qq\bar q \bar q) + \ldots$ respectively. 
If one wants to allow for mixing to, for example,  $q\bar q g$ or 
$qq\bar q \bar q$ states then these must be explicitly included in the 
calculation. Such coupled channel problems are rather difficult to solve.
Thus it is imperative to include as much of the physics of Fock sector
mixing as possible directly into the Hamiltonian. For example, the 
phenomenologically important hyperfine splittings can 
 be incorporated by iterating the $q\bar q g$ vertex, thereby adding an
effective transverse gluon exchange operator to the Hamiltonian. Using
the RGI Hamiltonian automatically achieves this goal.

The remainder of this paper is organized as follows. A  model based on the 
 Coulomb gauge QCD Hamiltonian is introduced
in Section II and its high energy behavior is analyzed. To do this we impose a
chiral invariant regulator and then 
calculate the RGI Hamiltonian.  The counter-term structure and 
possible perturbative and nonperturbative renormalization schemes are then 
discussed. In Section III we show how effective operators are constructed and
examine the ultraviolet behavior of the 
quark creation operator, the axial current, and the scalar quark density.  
Section IV presents
nonperturbative, one-loop, many-body calculations of
 the vacuum gap equation, the quark condensate, and the dynamical quark
mass. We summarize and conclude in Section V.

\section{The Effective Hamiltonian}

Our starting point will be the Coulomb gauge Hamiltonian of QCD. 
There are many reasons for choosing to work in the Coulomb gauge. For example,
as stated in the Introduction, it is expedient to have explicit $\bar q q$ 
interactions in the Hamiltonian to make a connection with quark model 
phenomenology. 
%A simple and effective first step in this 
%direction is to work in the Coulomb gauge.  
%Another the advantage of the Coulomb gauge is that 
%the confining interaction can easily be identified with the non-Abelian 
%Coulomb term in the Hamiltonian \cite{ss2}. EXPAND
Further advantages of the Coulomb gauge are that all unphysical degrees of 
freedom have been eliminated so that subsidiary 
conditions on states are not necessary. This also implies that the norm 
is positive definite. Furthermore, spurious retardation effects are minimized.

We note that it is impossible to address the issue of gauge invariance in this
model because it is defined in a specific gauge. However, QCD is gauge invariant
so it is perfectly acceptable to work in a fixed gauge. The real issue is not
how the model changes under a gauge transformation, but how well the 
phenomenological low energy interaction $V_{conf}$ we will use 
mimics the actual low energy behavior of QCD. The only way to address this is to
solve QCD (on the lattice, say) or to compare the predictions of the model
to observables. This will be the subject of future papers. For now, we note 
that
lattice gauge theory indicates that $V_{conf}$ should be a linear potential
when the quarks are static, our predicted glueball spectrum agrees well with
lattice calculations\cite{ssjc}, and the model is guaranteed to reproduce the 
success of the naive quark model for heavy quarks.

The DQM Hamiltonian is obtained after a rather long series 
of operations.  
The first step is to regulate the bare Hamiltonian in a manner which is consistent with the renormalization scheme we shall employ and which 
does not interfere with the nonperturbative calculations we will eventually perform. This introduces a scale into the Hamiltonian which we shall call
$\Lambda_0$. Renormalization is carried out by performing a similarity 
transformation on the Hamiltonian. This transformation is designed to 
eliminate couplings between the  scale $\Lambda_0$ and an arbitrary,
but nearby, lower scale $\Lambda_1$.
The similarity transformation preserves the spectrum of the Hamiltonian
(to the order we work to) while restricting 
 the matrix elements of the Hamiltonian to a smaller, band-diagonal form. 
 Since elimination of these couplings 
reveals UV divergences in the bare Hamiltonian    
the third step is to absorb these by fixing   
appropriate counter-terms. After this we repeat the last two steps above,
thereby generating a renormalization group improved effective Hamiltonian.
 The final step of our program is
to include a phenomenological term which is meant to model the behavior 
of the higher order terms that have been ignored in the perturbative 
 analysis. 
%A 
%side benefit of this is that it allows the strong coupling to remain 
%small -- even in the nonperturbative energy regime. 

\subsection{Model Definition}

As mentioned above, our starting point is  
the  
Coulomb gauge Hamiltonian of QCD \cite{TDL},

\begin{equation}
H = H_0 + V, \label{HTDL}
\end{equation}

\noindent
where $H_0$ is the free Hamiltonian, 

\begin{equation}
H_0 = \int d\x \psi(\x)^\dagger\left[-i\bbox{\alpha}\cdot \bbox{\nabla} + \beta m\right]\psi(\x)
 + Tr \int d\x \left[\bbox{\Pi}^2(\x) + {\bf B}_{\cal A}^2(\x)\right]. 
\label{hfree}
\end{equation}

\noindent
The degrees of freedom are the transverse gluon field $\A = \A^a T^a$,
its conjugate momentum  $\bbox{\Pi}$, and the quark field
in Coulomb gauge. We have represented the Abelian portion of the non-Abelian
magnetic field by ${\bf B}_{\cal A}$. The interactions are given by

\begin{eqnarray}
V &=& Tr\int d\x \left[ {\cal J}\bbox{\Pi}(\x) {\cal J}^{-1} 
\bbox{\Pi}(\x) - \bbox{\Pi}^2(\x)\right]
+ Tr\int d\x \left [\B^2(\x) - \B_{\cal A}^2(\x) \right] \nonumber \\
&+& {1\over 2} g^2\int d\x d{\bf y} \, {\cal J}^{-1} \rho^a(\x) 
\langle \x,a|(\bbox{\nabla}\cdot \D)^{-1} (-\bbox{\nabla}^2)
(\bbox{\nabla}\cdot \D)^{-1} | \y,b \rangle  {\cal J} \rho^b({\bf y}) \nonumber \\
&-& g \int d\x \psi^\dagger(\x) \bbox{\alpha}\cdot \A(\x) \psi(\x) \label{QCD}
\end{eqnarray}

\noindent
where ${\cal J} = \mbox{Det}[\bbox{\nabla}\cdot \D]$, 
$\D = \bbox{\nabla} - g \A$, is the covariant derivative in the adjoint 
representation and 
$\B = B_i = {\nabla}_j A_k - {\nabla}_k A_j  + g [\A_j,\A_k]$. 

The density, $\rho$, in the non-Abelian Coulomb interaction is the full 
 QCD color charge and thus 
 has both quark and  gluonic components,

\begin{equation}
\rho^a(\x) = \psi^{\dag}(\x) {\rm T}^a \psi(\x) + f^{abc} \A^b(\x) \cdot
\bbox{\Pi}^c(\x).
\end{equation}

\subsection{The Similarity Renormalization Scheme}

An effective Hamiltonian is defined to be an operator which, when acting within a subspace of the original Hilbert space, yields the same observables 
as the original Hamiltonian. 
Usually the effective Hamiltonian is constructed by dividing the Hilbert space 
into two portions which are defined by projection operators, $P$, and 
$Q=1-P$. 
In general the effective Hamiltonian may be written as a sum of 
the projected Hamiltonian,  $PHP$ and a set of effective interactions, 
$V_{eff}$ arising from elimination of couplings between the $P$ and $Q$ spaces.  
Typical methods for constructing the effective 
Hamiltonian produce energy denominators in $V_{eff}$ which may vanish. 
This
introduces severe complications when attempting to nonperturbatively 
diagonalize the effective Hamiltonian. An elegant way for avoiding these
problems is given by the similarity transformation scheme of G{\l}azek and
Wilson \cite{GW}. 

The scheme involves ordering the states according to their free energy levels.
The interaction part of the Hamiltonian is then regulated by ignoring 
matrix elements with free 
energy differences which are greater than $\Lambda_0$. One then proceeds
by constructing a similarity transformation which removes the couplings 
between states whose energy differences lie between $\Lambda_0$ and 
$\Lambda_1$. This procedure generates an effective potential which incorporates the eliminated physics. 

We start by constructing the eigenstates of the free Hamiltonian, $H_0$,
by expanding the quark and gluon field operators in normal modes,

\begin{equation}
\psi({\bf x})  = \sum_\tau \int {{d\k}\over {(2\pi)^3}}
\left( u(\k,\tau) b(\k,\tau) + v(-\k,\tau) d^{\dag}(-\k,\tau)
 \right)
{\rm e}^{i {\bf k}\cdot {\bf x}}
\end{equation}

\begin{equation}
\A({\bf x}) = \sum_a\int {{d\k}\over {(2\pi)^3}}
{1\over \sqrt{2 \omega{(\k)}}}
\left( \a(\k,a) +
 \a^\dagger(-\k,a) \right) {\rm e}^{i {\bf k}\cdot {\bf x}}.
\end{equation}

\noindent
The quark operators are given in the helicity basis and 
 all discrete quantum numbers    (helicity, color and flavor, 
 for the quarks and color for the gluons) 
 are collectively denoted by $\tau$ and $a$ respectively. 
Note the use of non-relativistic normalization so that, 
$u^{\dag}u=v^{\dag}v=1$, and 
\begin{equation}
\{b(\k_1,\tau_1), b^{\dag}(\k_2,\tau_2)\} = 
\{d(-\k_1,\tau_1),d^{\dag}(-\k_2,\tau_2)\} = (2\pi)^3\delta(\k_1-\k_2)\delta_{\tau_1,\tau_2}
\end{equation}
and
\begin{equation}
[\a(\k_1,a_1), \a^\dagger(\k_2,a_2)] =  (2 \pi)^3 \delta(\k_1 - \k_2)
\left({\bf 1} - \hat\k_1\hat\k_1 \right)\delta_{a_1a_2}.
\end{equation}

The free Hamiltonian is then given by, 

\begin{equation}
H_0  = \sum_\tau \int {{d\k}\over {(2\pi)^3}} E(\k)\left[b^{\dag}(\k,\tau)
b(\k,\tau) 
 + d^{\dag}(\k,\tau) d(\k,\tau) \right]
+ \sum_a \int {{d\k}\over {(2\pi)^3}} \omega(\k) 
 \a^{\dag}(\k,a)\a(\k,a) \label{hfree2}
\end{equation} 
where $E(\k) = \sqrt{m^2 + \k^2}$, $\omega(\k) = |\k|$ and we have 
dropped the constant, zero point energy. 
 
In the basis of eigenstates of $H_0$, $H_0 | n \rangle = E_n \vert n \rangle$, the cut-off method of G{\l}azek and Wilson may be written as  

\begin{equation}
\langle m|H|n \rangle \rightarrow \langle m|H^{\Lambda_0} |n \rangle  
= E_n \delta_{nm} + f_{mn}(\Lambda_0)
 \langle m |V| n \rangle. \label{barecut}
\end{equation}
Here $f_{mn}(\Lambda_0)$ is a function which is unity for 
$|E_{mn}| << \Lambda_0$  and zero for  $|E_{mn}| >> \Lambda_0$   
(we have defined $E_{mn} = E_m - E_n$).  
For our purposes we find it convenient to use a sharp cut-off,
\begin{equation}
f_{mn}(\Lambda_0) = \theta(\Lambda_0 - |E_{mn}|) \label{theta}.
\end{equation}

In this paper we will study only the quark sector of the 
Hamiltonian. Thus all gluon interactions will be ignored except those 
which contribute to effective quark operators.  
A discussion of the gluonic sector has been given in Ref. \cite{ssjc},
where it has been successfully applied to the glueball spectrum. The  
effort in deriving effective operators in the gluon sector 
is currently in progress \cite{ssjc2}. 

The operators of the bare regulated Hamiltonian contributing to order 
$g^2$ in the quark sector are,

\begin{equation}
V^{\Lambda_0} = g V^{\Lambda_0}_1 + g^2 V^{\Lambda_0}_2 
\end{equation}
where
\begin{eqnarray}
 V_1^{\Lambda_0} &=& \sum_{\tau_1\tau_2a}
\int {d \k_1\over (2 \pi)^3}{d \k_2\over (2 \pi)^3}
{d \bl\over (2 \pi)^3} 
\delta(\k_1 - \k_2 - \bl) \, { T^a_{12} \over \sqrt{2 \omega(\bl) }} \nonumber \\
&& \Bigg[ [u_1^\dagger {\bbox \alpha} u_2] \,
\theta(\Lambda_0 - | E(\k_2) + \omega(\bl) - E(\k_1)|)  \,
 b^{\dag}(\k_1,\tau_1) b(\k_2,\tau_2) \a(\bl,a) + \cdots \Bigg] 
\end{eqnarray}
and 

\begin{eqnarray}
V_2^{\Lambda_0} = \sum_{\tau_1\cdots \tau_4} &&
\int {d \k_1\over (2 \pi)^3}{d \k_2\over (2 \pi)^3}
{d \k_3\over (2 \pi)^3} {d \k_4\over (2 \pi)^3}
 V_c(\k_1 - \k_2) \, \delta(\k_1 + \k_2 - \k_3 - \k_4) \, 
T^a_{13} T^a_{24} \nonumber \\
&& \times \Bigg[ \theta(\Lambda_0 - 
 | E(\k_3) + E(\k_4) - E(\k_1) - E(\k_2)|) \,
[-u_1^\dagger u_3] [u_2^\dagger u_4] \nonumber \\ 
 && b^{\dag}(\k_1,\tau_1)  b^\dagger(\k_2,\tau_2)  b(\k_3,\tau_3) 
b(\k_4,\tau_4)  + \cdots \Bigg].
\end{eqnarray}
In $V_1$ and $V_2$ the ellipses denote an additional seven and fifteen terms 
respectively and $V_c$ is the momentum space Coulomb potential. 
 The additional terms correspond to interactions involving all 
possible permutations of 
quark and antiquark operators. The subscripts on the spinors and the Gell-Mann
matrices refer to indices of the appropriate quark operators. 
Normal ordering the bare Hamiltonian with respect to the 
perturbative basis generates a new one body operator
 given by
\begin{eqnarray}
g^2 V_{self.C} = \sum_{\tau} \int { {d\k} \over {(2\pi)^3}} &&
\Big[
 \Sigma_C(\k;\Lambda^C_{0}) 
[b^{\dag}(\k,\tau) 
b(\k,\tau) + 
 d^{\dag}(-\k,\tau) d(-\k,\tau) ]  \nonumber \\
&& + \theta( {\Lambda_0 \over 2} - E(\k))
G_C(\k;\Lambda^C_{0}) 
[b^{\dag}(\k,\tau) 
d^{\dag}(-\k,\tau) + 
 d(-\k,\tau) b(\k,\tau) ] \Big].
\label{vc}
\end{eqnarray}
Both $\Sigma_C$ and $G_C$ (which are given in the Appendix)
are divergent and not regulated by $\Lambda_0$. In Eq.~[\ref{vc}] they have 
have been regulated individually with a cutoff denoted by $\Lambda^C_{0}$.
Since all divergences of the bare Hamiltonian are to be 
absorbed by counter-terms one could subtract this 
self energy in its entirety and simply ignore it. However
  we shall show subsequently that 
covariance requires that a portion of $V_{self.C}$ be maintained in the 
Hamiltonian.  
The $O(g^2)$ bare potential will therefore be given by,  

\begin{equation}
V^{\Lambda_0}(\Lambda^C_0) =  g V^{\Lambda_0}_1 + g^2 V^{\Lambda_0}_2 + g^2 
V^{\Lambda_0}_{self.C}(\Lambda^C_0)
\end{equation}
where we have made the $\Lambda_0$ and $\Lambda_C^0$ dependence explicit.
The $\Lambda_0$ dependence of $V_{self.C}$ arises only from the 
term proportional to $G_C$ since $\Sigma_C$ is 
diagonal in the perturbative basis and therefore, according to 
Eq.~[\ref{barecut}], is $\Lambda_0$ independent.
So long as the ratio 
$\Lambda^C_0/\Lambda_0$ remains finite as both scales approach infinity,
the existence of a second scale will be irrelevant (to the order to which 
we work). From now on we will set $\Lambda^C_0 = \Lambda_0$.

We are now ready to calculate the equivalent of the $P$-space 
Hamiltonian, $PHP=H^{\Lambda_1}$ and of the 
effective Hamiltonian, $H^{\Lambda_1}_{eff}$. 
The matrix elements of the former  
are defined using Eq.~[\ref{barecut}],

\begin{equation}
\langle m|H^{\Lambda_1}|n \rangle 
= E_n \delta_{nm} + f_{mn}(\Lambda_1)
 \langle m |V| n \rangle \label{reg}, 
\end{equation}
where $V$ depends on $\Lambda_0$ (and $\Lambda^C_0$ in general) 
as discussed above. 
Since within the $P$-space the spectrum of the bare Hamiltonian   
and that of the effective Hamiltonian are to be identical, we may write
\begin{equation}
H^{\Lambda_1}_{eff} = S H^{\Lambda_0} S^{-1}
\end{equation}
where $S$ is a unitary matrix such that 
matrix elements, $\langle m|H^{\Lambda_1}_{eff}|n \rangle$ 
are nonzero only within the $P$-space, i.e. for states 
that satisfy 

\begin{equation}
|E_{nm}| < \Lambda_1. \label{below}
\end{equation}

To construct $S$ we proceed by expanding it in powers of the strong 
coupling

\begin{equation}
S = {\rm e}^{i R},
\end{equation}
and
\begin{equation}
R = g R_1 + g^2 R_2 + {\cal O}(g^3).
\end{equation}
Then

\begin{eqnarray}
& & H^{\Lambda_1}_{eff} =  \phantom{+} 
 H_0 + g\left(V_1 + i  [R_1,H_0]\right)  \nonumber \\
& & \phantom{H^\Lambda_{eff} = }
+ g^2 \left( V_2 + V_{self.C} 
+ i [ R_2, H_0] + i  [R_1,  V_1] - {1\over 2}  
 [R_1,[R_1,H_0]] \right) + O(g^3) \label{sim2}.
\end{eqnarray}
Here, the dependence of the $V$'s on $\Lambda_0$ is implicit. 
To order $g$, $ig [R_1,H_0]$ is chosen to eliminate matrix elements of 
$g V_1$ between the scales $\Lambda_1$ and $\Lambda_0$. Thus, to this order, 
one
need simply replace the two terms in the first brackets with $V_1^{\Lambda_1}$,
\begin{equation}
\langle m|V_1^{\Lambda_1}|n \rangle = \theta(\Lambda_1 - |E_{mn}|)
 \langle m|V_1|n \rangle.
\end{equation}
Similarly, to order
$g^2$, $i g^2[R_2, H_0]$ eliminates all matrix elements of the expression 
in the second brackets 
between $\Lambda_0$ and $\Lambda_1$.
A new interaction in $H^{\Lambda_1}_{eff}$ which distinguishes it from 
$H^{\Lambda_1}$ arises from the double commutator term in 
Eq.~[\ref{sim2}] when it is evaluated for matrix 
elements satisfying Eq.~[\ref{below}]. 
 $H^{\Lambda_1}_{eff}$ is thus given by 

\begin{equation}
H^{\Lambda_1}_{eff} = H_0 + gV_1^{\Lambda_1} + g^2 V_2^{\Lambda_1} 
+ g^2 V^{\Lambda_1}_{self.C} 
 + g^2 V^{\Lambda_1}_T + O(g^3). \label{heff}
\end{equation}
where
\begin{equation}
\langle m|
 V_2^{\Lambda_1} 
+ V^{\Lambda_1}_{self.C} |n \rangle = \theta(\Lambda_1 - |E_{mn}|)
\langle m|
 V_2
+ V_{self.C} |n \rangle . 
\end{equation}

\noindent
The last term in Eq. [24] has been generated by the similarity transformation and is  meant to reproduce the physics lost when the cutoff is reduced from
$\Lambda_0$ to $\Lambda_1$.
For sharp cutoffs (see Eq.~[\ref{theta}]) one has
\noindent
\begin{equation}
\langle m| V^{\Lambda_1}_T|n \rangle  = \theta(\Lambda_1 - |E_{mn}|)
\sum_q \langle m \vert V_1 \vert q \rangle \langle q
 \vert V_1 \vert n\rangle \Theta_T(E_{qn},E_{qm};\Lambda_1). 
\end{equation}
Note that the sum over intermediate states is cut-off from above 
due to an implicit dependence of the matrix elements of $V_1$ on $\Lambda_0$.
The last factor is a function which arises in the similarity transformation
and is similar to an energy denominator in standard perturbation theory. It is
given by
\begin{equation}
\Theta_T(\Delta_a,\Delta_b;\Lambda) = -{{\theta( |\Delta_a| - \Lambda)} 
\over {\Delta_a}} \left(1 - {1\over 2}\theta( |\Delta_b| - \Lambda) 
\right) + (a \leftrightarrow b) \label{theta_T}
\end{equation}
Notice, that energy denominators never cause divergences 
since they are cut off from below by $\Lambda_1$.

We are now in a position to study the ultraviolet behavior of the theory. As $\Lambda_0$ approaches infinity
$V^{\Lambda_1}_2$ is finite since it is cutoff at the lower scale $\Lambda_1$.  Alternatively, the self energy $V_{self.C}$ diverges  as  
$\Lambda^C_0=\Lambda_0 \to \infty$. Similarly, the effective 
interaction $V^{\Lambda_1}_T$ contains self energy pieces which diverge and interaction pieces which are ultraviolet finite. Thus it is convenient to
split it into two portions

\begin{equation}
V^{\Lambda_1}_T = V^{\Lambda_1}_{self.T} + V^{\Lambda_1}_{ex.T},
\end{equation}
where 

\begin{eqnarray}
g^2 V^{\Lambda_1}_{self.T} &=& \sum_{\tau} 
\int {d {\bf k}\over (2 \pi)^3} \,
\Sigma_T(\k;\Lambda_0,\Lambda_1) 
[b^{\dag}(\k,\tau)  b^\dagger(\k,\tau) + d^{\dag}(-\k,\tau)d(-\k,\tau) ] 
\nonumber \\
&+& \theta( {\Lambda_1 \over 2} - E(\k) )
G_T(\k;\Lambda_0,\Lambda_1) [b^{\dag}(\k,\tau)d^{\dag}(-\k,\tau) 
+ d(-\k,\tau)b(\k,\tau) ]
\end{eqnarray}
and $V_{ex.T}$ has a similar structure to $V_2$. 
The functions $\Sigma_T$ and $G_T$ are given in the Appendix.
As stated above, in the limit $\Lambda_0 \to \infty$, $V^{\Lambda_1}_{self.T}$ is divergent
while the exchange interaction, $V^{\Lambda_1}_{ex.T}$ remains finite.

The divergences in the self energy pieces must be absorbed by adding 
appropriate counter-terms to the bare Hamiltonian. Thus we define

\begin{equation}
H = \lim_{\Lambda_0 \to \infty} \left[ H^{\Lambda_0}
 + g^2 V^{\Lambda_0}_{ct.} \right].
\end{equation}
Since only one-body operators are explicitly UV divergent 
we construct counter-terms of the form
\begin{eqnarray}
 g^2  V^{\Lambda_0}_{ct.} &=& \sum_{\tau}
\int {d {\bf k}\over (2 \pi)^3} \,
 \Sigma^{ct.}(\k;\Lambda_0)
[b^{\dag}(\k,\tau)b(\k,\tau) + d^{\dag}(-\k,\tau)d(-\k,\tau) ] \nonumber \\
& & \phantom{
  \sum_{\tau}
\int {d {\bf k}\over (2 \pi)^3} }
 +  \theta( {\Lambda_0 \over 2} - E(\k) )
G^{ct.}(\k;\Lambda_0)
[b^{\dag}(\k,\tau)d^{\dag}(-\k,\tau) + d(-\k,\tau)b(\k,\tau) ]. \label{hct}
\end{eqnarray}
Note that $V_{ct}^{\Lambda_0}$
 contains a piece which is off-diagonal in the perturbative basis ($G^{ct.}$).
This will lead to a contribution from $V_{ct.}^{\Lambda_0}$ in $R_2$ when $H^{\Lambda_0} 
 + g^2 V_{ct}^{\Lambda_0}$ is used as an input to the similarity 
transformation (as it should be).
This extra contribution to $R_2$ removes couplings caused by $G^{ct.}$  
when $\Lambda_1 < 2E(\k) < \Lambda_0$  and therefore only leaves
a nonvanishing contribution from this counter-term for
$\Lambda_1 > 2E(\k)$. 

\subsection{The Renormalization Group Improved Hamiltonian}

Our goal is to define an effective Hamiltonian which is operative at 
hadronic scales. Thus one may be tempted to set $\Lambda_1 \sim 1$ fm$^{-1}$. 
However, a single-step similarity transformation which evolves the 
Hamiltonian from the UV cut-off $\Lambda_0$ down to the hadronic scale 
would introduce effective interactions with large matrix elements 
 for which the perturbative approach is hard to justify. 
 This would also result in a large sensitivity of the effective Hamiltonian to 
 parameters which are set at the UV scale.  
  Instead we may use 
 the renormalization transformation iteratively over many small steps 
 and at each step adjust the Hamiltonian only by a small amount.
 In the language of standard Feynman perturbation theory, 
 the two approaches correspond to keeping only the leading-log 
 and summing the leading-log to all orders, respectively. 
 Thus we choose to apply the similarity transformation 
\begin{equation}
N = {{\log {\Lambda_0\over \Lambda_N}  }\over {\log(1 + \epsilon)} }
\end{equation}
times starting at the beginning of the renormalization group trajectory 
with the bare Hamiltonian, 
$H^{\Lambda_0}=H^{\Lambda_0}_{eff}$ and 
finishing at the end of the trajectory with $H^{\Lambda_N}_{eff}$.
Each time the transformation is applied the cutoff is reduced 
from $\Lambda_i$ to $\Lambda_{i+1}$ by a finite amount
\begin{equation}
 {\Lambda_{i} \over {\Lambda_{i+1}}} = 1 + \epsilon . \label{epsilon}
\end{equation}
The  limit $\Lambda_0 \to \infty$ corresponds to  
$N\epsilon$ diverging logarithmically.

Expanding the integrands in the expressions for $V_{self.C}$ and $V_{self.T}$ 
(Eqs. [15] and [29]) in powers of $\epsilon$ yields (see Appendix),
\begin{eqnarray}
\Sigma_T(\k;\Lambda_{0},\Lambda_1)  & \to & \epsilon {{g^2C_F}\over {(4\pi)^2}}
\left[ {{2m^2} \over {E(\k)}} - {{8\k^2}\over {3E(\k)}}
\right]\left(1 + O\left({1\over {e^{N\epsilon} }}\right) \right)  \nonumber \\
  \Sigma_C(\k;\Lambda_{0})  
& \to & \epsilon {{g^2C_F}\over {(4\pi)^2}}
\left[ {{4m^2} \over {E(\k)}} + {{8\k^2}\over {3E(\k)}}
\right]
\left(1 + O\left({1\over {e^{N\epsilon} }}\right) \right)
+ \Sigma_C(\k;\Lambda_1) \nonumber \\
G_T(\k;\Lambda_{0},\Lambda_1) & \to &  -\epsilon{{g^2C_F}\over {(4\pi)^2}}
 {{14|\k|m} \over {3E(\k)}}\theta
\left({\Lambda_1\over 2} - E(\k)\right)
\left(1 + O\left({1\over {e^{N\epsilon} }}\right) \right)  \nonumber \\
 G_C(\k;\Lambda_0) \to
 G_C(\k;\Lambda_{0},\Lambda_1)& \to &  -\epsilon {{g^2C_F}\over {(4\pi)^2}}
 {{4|\k|m} \over {3E(\k)}}
\theta\left({{\Lambda_1}\over 2} - E(\k)\right)
\left(1 + O\left({1\over {e^{N\epsilon} }}\right) \right) \nonumber \\
& \phantom{ \to} &  +  G_C(\k;\Lambda_1,\Lambda_1) .
\label{selfexp}
\end{eqnarray}

\noindent
The presence of the last terms in the second and fourth equations 
above causes a complication. Since $\Lambda_N$ is fixed at the end of 
 the trajectory these terms 
 are of $O(N\epsilon)$ and dominate transverse self energies 
 generated by the similarity transformation, which are of $O(\epsilon)$
(the first and third equations of Eq.~[\ref{selfexp}]).  
 In order for these two types of self energies to be comparable,
 most of the Coulomb  
 self energy induced by normal ordering must be subtracted 
 by a counter-term in the original Hamiltonian. 
An unusual feature of this is that the subtraction must depend on the 
renormalization group trajectory (as opposed to standard perturbation 
theory).  In other words it is not enough to fix the Coulomb counter-term  in
$H^{\Lambda_0}$, it must also depend on the trajectory ({\it i.e.}, it must 
be $\epsilon$-dependent).
This is to be expected: covariance relates 
both Coulomb and transverse gluon exchange and since the similarity 
transformation affects only the latter,
Coulomb self-energies must be adjusted in an appropriate fashion along the entire trajectory. 

For a Hamiltonian evaluated at $\Lambda_i$ along a trajectory defined by 
Eq.~[\ref{epsilon}] the counter-term is given by 
\begin{equation}
\Sigma^{ct.}(\k,\Lambda_i)  
 \to \Sigma^{ct.}(\k,\Lambda_i;\epsilon)   = 
   - \Sigma_C(\k,{{\Lambda_i}\over {1+\epsilon}}) \label{cct}
\end{equation}
and
\begin{equation}
G^{ct.}(\k,\Lambda_i)  \to  G^{ct.}(\k,\Lambda_i,\epsilon)
= - G_C(\k;{{\Lambda_i}\over {1+\epsilon}},\Lambda_{i-1}).
\end{equation}
The first portion of each equation is meant to indicate that the 
counter-terms have become trajectory dependent.
Possible finite pieces have not been considered in these expressions 
({\it i.e.}, scheme dependence) since they become irrelevant upon use of 
the renormalization group equations. 

At this stage the perturbative portion of the Hamiltonian is fully 
specified and we are ready to apply the renormalization group procedure. 
After a single step ($\Lambda_0 \to \Lambda_1$) a new one-body operator
is generated,

\begin{eqnarray}
& & \int {{d\k} \over {(2\pi)^3}} 
 \left[\sqrt{\k^2+m^2}  + \Sigma(\k;\Lambda_0,\epsilon) 
    \right] 
[b^{\dag}(\k,\tau) b(\k,\tau) +
d^{\dag}(-\k,\tau) d(-\k,\tau)] 
\nonumber \\  
& &  + \int {{d\k} \over {(2\pi)^3}}  \theta({\Lambda_1\over 2}-E(\k)) 
G(\k;\Lambda_0,\epsilon) 
 [ b^{\dag}(\k,\tau) d^{\dag}(-\k,\tau) + 
d(-\k,\tau) b(\k,\tau)]. 
\label{onebody2} 
\end{eqnarray}
We have defined new quantities

\begin{equation}
\Sigma(\k;\Lambda_0,\epsilon) = \Sigma_T(\k;\Lambda_0,\Lambda_1)
 + \Sigma_C(\k,\Lambda_0) + \Sigma^{ct.}(\k,\Lambda_0,\epsilon) 
 = \epsilon {{g^2C_F}\over {(4\pi)^2}}
{{6m^2} \over {E(\k)}} 
\left(1 + O\left({1\over {e^{N\epsilon} }}\right) \right) 
\end{equation}
and

\begin{equation}
G(\k;\Lambda_0,\epsilon) = G_T(\k;\Lambda_0,\Lambda_1)
 + G_C(\k,\Lambda_0) + G^{ct.}(\k,\Lambda_0,\epsilon) 
 = -\epsilon {{g^2C_F}\over {(4\pi)^2}}
{{6|\k|m} \over {E(\k)}} 
\left(1 + O\left({1\over {e^{N\epsilon} }}\right) \right). 
\end{equation}
The last equality in each equation follows from Eqs.~[34], [35], and [36].
In order to iterate the similarity transformation further the one body 
operator (Eq. [37]) must be diagonalized. This is because the eigenstates
of this operator must be used as a new basis in which matrix elements of
the interaction are regulated.
We therefore proceed by
rotating the quarks back to the massless basis 
($b,d \to \hat b,\hat d$).
%\begin{eqnarray}
% \hat b(\k,\tau) & = & c_\h(\k) b(\k,\tau) + s_\h(\k) 
% d^{\dag}(-\k,\tau) \nonumber \\
% \hat d(-\k,\tau) & = &  c_\h(\k) d(-\k,\tau) - s_\h(\k) b^{\dag}(\k,\tau)
%\label{BV0}
%\end{eqnarray}
%with  $c_\h = \cos(\h\phi_m(\k))$ and $s_\h = \sin(\h\phi_m(\k))$ and
%\begin{equation}
%s(\k) = {m\over {\sqrt{m^2 + \k^2}}},\;\; c(\k) 
%= {{|\k|}\over \sqrt{m^2 + \k^2}}. \label{BVm}
%\end{equation}
Eq. ~[\ref{onebody2}] then becomes 

\begin{eqnarray}
& & \int {{d\k} \over {(2\pi)^3}} 
 \left[|\k| + {\cal O}({g^2\over {e^{N\epsilon}}})\right]  
[\hat b^{\dag}(\k,\tau) \hat b(\k,\tau) + 
\hat d^{\dag}(-\k,\tau) \hat d(-\k,\tau)] 
\nonumber \\  
& &  - m_1\int {{d\k} \over {(2\pi)^3}}  \theta({\Lambda_1\over 2}-E(\k)) 
[ \hat b^{\dag}(\k,\tau) \hat d^{\dag}(-\k,\tau) + 
\hat d(-\k,\tau) \hat b(\k,\tau)] 
\nonumber \\
\label {Zs}
\end{eqnarray}
where
\begin{equation}
m_1 = m\left[ 1 + g^2 {{6C_F} \over {(4\pi)^2}}
\epsilon\left( 1 + {\cal O}({1\over {e^{N\epsilon}}}) \right) \right]
\end{equation}
Note that it is only necessary to rotate the basis below the scale $\Lambda_1$
so that the $\theta$-functions may be ignored in this process.

To order $g^2$ this looks just like the free Hamiltonian 
 with $m$ replaced by $m_1$ and quantized in a 
massless basis. We now rotate the one body operator back to the
massive quark basis using $m_1$ as the bare quark mass and obtain a free, 
 diagonal Hamiltonian, $H_{eff}^{\Lambda_1}(m_1)$ which 
 may be used as input to the next similarity transformation ($\Lambda_1 \to
\Lambda_2$). Note that since the similarity transformation is only 
affected to order $g^2$ by these rotations, the net effect of this 
procedure is simply to replace $m_0 = m(\Lambda_0)$ by $m_1 = m(\Lambda_1)$. 
The remainder of the renormalization group procedure is simple: we apply the above transformations $N$ times and obtain an  
effective Hamiltonian with the one-body operator given by the free 
Hamiltonian with the replacement $m_0 \to m_N = m(\Lambda_N)$:

\begin{equation}
m_0 \to m_N = 
m\left[ 1 + g^2 {{6C_F} \over {(4\pi)^2}}
\epsilon\left( 1 + O({1\over {e^{N\epsilon}}}) \right) \right]^N.
\end{equation}
In the limit $N\epsilon \sim \log{\Lambda_0/\Lambda_N} \to \infty$ 
this becomes

\begin{equation}
m(\Lambda_N) = m(\Lambda_0)\left( {\Lambda_0 \over \Lambda_N} 
\right)^{  g^2 {{6C_F} \over {(4\pi)^2}}}. \label{massrun}
\end{equation}
\noindent

Finally, the remainder of the RGI Hamiltonian (the two-body terms) is 
given by the interactions defined at the end scale 
 
\begin{equation}
g V_1^{\Lambda_N} + g^2 V_2^{\Lambda_N} + g^2 V_{ex.T}^{\Lambda_N}.
\end{equation}

It is perhaps useful to summarize the calculation to this point.
We have started with the exact Coulomb gauge Hamiltonian of QCD, expanded
it to order $g^2$, and regulated it in a manner which lends itself to
the similarity transformation renormalization procedure. Performing the
transformation generated Coulomb and transverse gluon exchange self
energy effective interactions which diverge in the ultraviolet limit.
The divergences were absorbed into counter-terms and the whole process
was iterated to construct the RGI Hamiltonian. This Hamiltonian is
defined at a hadronic scale $\Lambda_N$, is ultraviolet and infrared
finite, and has nonzero matrix elements only for states which differ
in energy by up to $\Lambda_N$ 
(higher energy physics is incorporated in the effective interaction, $V_{ex.T}$). Notice that all self energy interactions have been absorbed
by the renormalization procedure and no longer explicitly appear in the
Hamiltonian. As expected for the Coulomb gauge, there is no need for 
 wave function renormalization and the coupling constant does not run until 
 $O(g^3)$.

%In QCD perturbative evolution of Hamiltonian has to be stopped when 
% $\Lambda \gtrsim 1\mbox{fm}^{-1}$ i.e. when the coupling constant 
% becomes large. 
% At his point the effective Hamiltonian 
% has to be solved nonperturbatively. 
% $\Lambda$-independence of physical observables including the   
% spectrum of $H^{\Lambda}_{eff}$ is guaranteed by the  nonperturbative 
%renormalization group trajectories of the parameters in the effective 
%Hamiltonian. For large $\Lambda$ these trajectories will match those 
%   computed using perturbative approximation to the similarity 
% transformations e.g. Eq.~[\ref{massrun}]. 

\subsection{The Phenomenological Interaction}

The RGI Hamiltonian which we have constructed can be reliably employed to
calculate observables down to a scale of $\Lambda \sim 1 {\rm fm}^{-1}$
where the coupling constant becomes large (note that we shall refer to the
end point of the renormalization group trajectory as $\Lambda$ from now on). 
At this point the Hamiltonian 
must both be solved nonperturbatively and must be supplemented with a 
phenomenological potential which incorporates as much of the neglected 
higher order physics as possible. The latter forms the topic of this
section.

Choosing an effective low energy interaction is not trivial because the
model is relativistic. This means that a Dirac structure must be imposed
on the interaction -- with commensurate implications on phenomenology. 
It is commonly held that the effective potential should be linear with a
scalar current interaction. 
 However, we have shown that this simple expectation is
misleading \cite{ss2}. One may establish the Dirac
nature of the low energy effective potential in the heavy quark limit.
 In this case it is appropriate to make a nonrelativistic reduction and 
 use the quenched approximation.  $P$-wave splittings in the $J/\psi$ and 
$\Upsilon$ spectra imply that long range spin-dependent potentials 
correspond to a nonrelativistic reduction of 
 a scalar, phenomenological, confining interaction~\cite{schnitzer}. 
 This view is also supported by 
 lattice gauge calculations\cite{schnitzer} and phenomenological, minimal-area-law model 
 for the Wilson loop~\cite{Nora}.
 Starting with the Coulomb gauge Hamiltonian, 
 it is possible to show that the effective 
 scalar interaction arises dynamically through mixing with hybrids 
 while the static part of the $\bar q q$ potential remains of a 
timelike-vector nature~\cite{ss2}.

If one assumes that the heavy quark potential also applies
for light quarks then the net result is to simply replace the order $g^2$
Coulomb potential with a Coulomb plus linear potential:

\begin{equation}
-g^2 C_F{1\over {4\pi r}} \rightarrow -g^2 C_F{1\over {4\pi r}} 
 + \sigma r. \label{modify}
\end{equation}

\noindent
We shall fix the string tension at $\sigma = 0.18$ GeV$^2$, commensurate with
Regge phenomenology, lattice gauge theory, and the constituent quark model.
 Note that the phenomenological two-body potential is meant to reproduce 
 low energy physics, thus Eq.~[\ref{modify}]
 is used in the effective Hamiltonian, {\it i.e.},  below the scale $\Lambda$. 
 We also introduce a one-body quark operator which corresponds to the linear
 term in Eq. [45]. 
This eliminates color non-singlet states from the spectrum. Furthermore,
it ensures that color-singlet bound states are infrared-finite.

These observations will be 
discussed in greater detail below.
The final form for the quark sector of the model Hamiltonian which 
we shall use in subsequent calculations is:

\begin{equation}
H_{DQM} = H_0(\Lambda) + gV^\Lambda_1 + g^2 V^\Lambda_2 
 + g^2 V^\Lambda_{ex.T} + V_{conf} + V_{self.conf}.
\end{equation}

\noindent
This has been defined at the scale $\Lambda$; 
hence the interaction terms are all cutoff at this scale. As discussed above, 
the $\Lambda$ dependence of $H_0$ comes from the running mass. Finally 
$V_{conf}$ has only two-body matrix elements.

\section{Effective Operators}

If one wishes to make statements about matrix elements which are accurate to
order $g^2$ it is not sufficient to know the states 
({\it i.e.}, the Hamiltonian) 
to this order -- one must also evolve the relevant operator. 
This is illustrated below with explicit calculations of the 
perturbative evolution of 
the single quark operator, the axial charge operator, and 
the composite operator $\bar q q$ 
used to define the chiral quark condensate. 

Just as with the Hamiltonian, one may use the similarity 
renormalization 
scheme to calculate effective operators  restricted to 
$P$-space.
Following the steps discussed in Sec.~II we first define an operator 
 ${\cal O}^{\Lambda_0}$ by restricting its matrix elements in the basis of 
eigenstates of $H_0$,

\begin{equation}
\langle m |{\cal O} |n \rangle \to 
\langle m |{\cal O}^{\Lambda_0}| n \rangle 
 = \theta(\Lambda_0-|E_{mn}|) \langle m |{\cal O} |n \rangle \label{oevol}
\end{equation}
The effective 
operator is then given by

\begin{equation}
{\cal O}_{eff} = S {\cal O} S^{-1} = {\cal O} + i g [R_1, {\cal O}] + g^2
 \left(
[R_2, {\cal O}] - {1\over 2}[R_1,[R_1,{\cal O}]] \right) + O(g^3)
\label{eq:Oeff}
\end{equation}

Notice that in contrast to the evaluation of the Hamiltonian, explicit 
expressions for 
$R_1$ and $R_2$ are required to compute one-loop corrections to operators.
In terms of eigenstates of $H_0$ these are

\begin{equation}
\langle m|R_1|n \rangle  = i \theta(\Lambda_1 - |E_{mn}|)
 {{ \langle m| V_1|n \rangle } \over E_{nm}}
\end{equation}

\noindent
and

\begin{equation}
\langle m|R_2| n\rangle  = i {{\theta(\Lambda_1 - |E_{nm}|)} \over 
E_{nm}} 
 \left( \langle m|V_2|n \rangle + 
\sum_q \langle m|V_1|q\rangle\langle q|V_1| n\rangle 
 \Theta_T(E_{qn},E_{qm};\Lambda_1) \right).
\label{eq:R2}
\end{equation}
Calculating radiative corrections to operators thus becomes a 
straightforward, albeit tedious, process of evaluating commutators.
Note that the similarity transformation may be iterated to form the 
renormalization group improved effective operator, just as for the 
Hamiltonian.
  
Matrix elements of an operator ${\cal O}$, 
\begin{equation}
\langle m|{\cal O}|n \rangle = \langle m| S^{-1} {\cal O}_{eff} S |n \rangle
\end{equation}
have mixings to eigenstates with large energy included perturbatively in 
 ${\cal O}_{eff}$ 
 This enables one to resolve an
ambiguity in applying perturbative QCD to hadronic matrix
elements. The ambiguity is illustrated in Fig. 1 where two possible diagrams
contributing to the hadronic matrix element of a generic current are presented.
The question is if the current couples to nonvalence structure in the hadron
(say, the strangeness content of the proton), does it do so via an existing 
nonvalence component of the wave function (as in Fig. 1a) or via a virtual 
quark loop (as in Fig. 1b)? It is impossible to resolve this in typical 
approaches to this problem ({\it i.e.}, without reference to dynamics). 
The method presented here allows one to identify both contributions. 
  Fig. 1a follows from solving the bound state effective Hamiltonian in a 
basis which allows the appropriate Fock state mixing. Thus the intermediate
states associated with the wave function exist {\it below} $\Lambda$. 
Alternatively, Fig. 1b arises from transforming the current ${\cal O} \to {\cal O}_{eff}$ and involves 
intermediate states which exist 
{\it above} $\Lambda$. We emphasize that correctly interpreting
this situation can only be done in the context of a consistent dynamical 
model. 

\hbox to \hsize{%
\begin{minipage}[t]{\hsize}
\begin{figure}
\epsfxsize=5in
\hbox to \hsize{\hss\epsffile{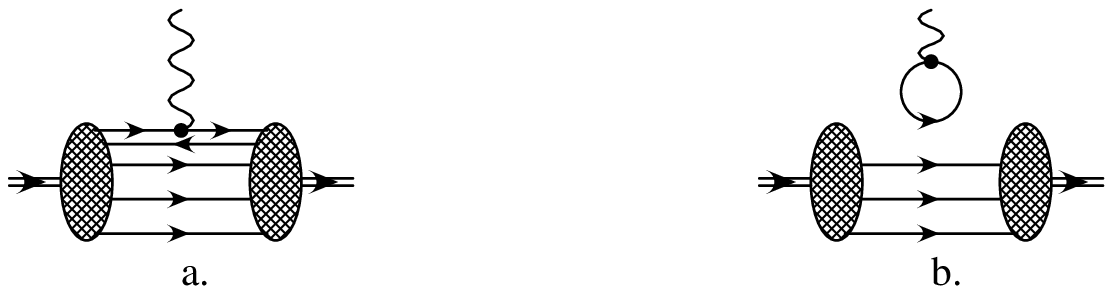}\hss}
\label{fig:amb}
\end{figure}
\end{minipage}}
\begin{center}
  {\small Fig.~1. Resolving Ambiguities in Current Matrix Elements.}
\end{center}

\subsection{The Effective Quark Operator}

As a first simple application of the similarity transformation we evaluate
the one-loop correction to the bare quark operator, $b({\bf k},\tau)$. This
means evaluating the commutators indicated in Eq.~[\ref{eq:Oeff}] with 
${\cal O} = b({\bf k},\tau)$. Since 
we are interested in the evolution of the operator itself (and not the 
generation
of new operators such as $b a^\dagger$), the first order commutator 
($[R_1, b({\bf k},\tau)]$) does not contribute. Furthermore the order $g^2$ 
commutator, $[R_2, b({\bf k},\tau)]$, is zero. This is because the first term 
in $R_2$ (see Eq.~[\ref{eq:R2}]) involves the Coulomb potential which is 
normal ordered. Thus all commutators with the quark annihilation operator
simply yield terms which are proportional to the Coulomb interaction 
(and never
$b({\bf k},\tau)$). The commutators of the second term in  $R_2$ cancel in
pairs and again, do not contribute.  Thus the only contribution to $b_{eff}$
is the double commutator over $R_1$ in Eq.~[\ref{eq:Oeff}]. Two of the
64 possible terms are nonzero, these are shown diagrammatically in Fig.~2a,b.
As Fig. 2 indicates, the process of evaluating commutators for ${\cal O}_{eff}$
is similar to performing time-ordered perturbation theory. However, we warn the
reader that this is an analogy only since nontrivial cut-off dependence is
implicit in each diagram.

\hbox to \hsize{%
\begin{minipage}[t]{\hsize}
\begin{figure}
\epsfxsize=5in
\hbox to \hsize{\hss\epsffile{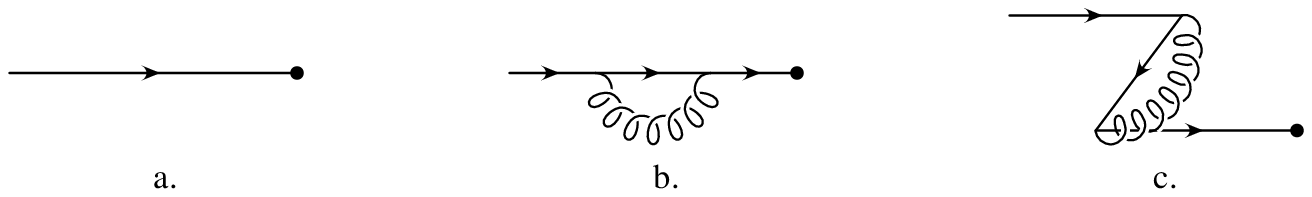}\hss}
\label{fig:beff}
\end{figure}
\end{minipage}}
\begin{center}
  {\small Fig.~2. Diagrams Contributing to $b_{eff}$.}
\end{center}

Expressions for these contributions to $b_{eff}$ are respectively

\begin{equation}
- g^2 {C_F \over 4} 
\int {d\p \over (2 \pi)^3} {(1 - \hat\p\cdot\hat\bl \, 
\hat\k\cdot\hat\bl) 
\over \omega(\bl)} { 
\Theta(\Lambda_0 > | E(\p) + \omega(\bl) - E(\k)| > \Lambda_1)
 \over (E(\p) + \omega(\bl) - E(\k))^2}
\end{equation}

\begin{equation}
- g^2 {C_F \over 4} \int {d\p \over (2 \pi)^3} {(1 + \hat\p\cdot\hat\bl \, 
\hat\k\cdot\hat\bl) ) \over \omega(\bl)} 
{\Theta(\Lambda_0 > | E(\p) + \omega(\bl) + E(\k)| > \Lambda_1) \over
(E(\p) + \omega(\bl) + E(\k))^2},
\end{equation}

\noindent
here we have taken $\bl = \p - \k$ and 
$\Theta(a > b > c)=\theta(b-c)\theta(a-b)$. 
Applying the similarity transformation $N\sim \log{\Lambda_0/\Lambda_N}$ 
times to reduce the 
cut-off from $\Lambda_0$ to $\Lambda_N$ yields the following leading 
order result

\begin{equation}
b(\k,\tau) \to b(\k,\tau)\left( {\Lambda_0 \over \Lambda_N} 
\right)^{ -C_F {{g^2}\over {(4\pi)^2}}}
\end{equation}

When the similarity scale equals the ultraviolet
cutoff, no renormalization of the quark operator occurs as expected. As the 
similarity scale is reduced, the ``strength" of the quark operator diminishes
as ``non-diagonal" operators (such as $b a^\dagger$) become 
important.

\subsection{The Axial Charge at One-loop}

As a simple application of the renormalization of composite operators, we now
apply the methodology to the axial charge operator for massless quarks.
In terms of bare quark operators this is given as

\begin{equation}
Q_5 = \sum_\tau \int {d \k \over (2 \pi)^3} 
{\bf \sigma} \cdot \hat k [b^\dagger({\bf k},\tau) b({\bf k},\tau) + 
d^\dagger({\bf k},\tau) d({\bf k},\tau)].
\end{equation}

In this case 
the first six diagrams of Fig. 3 contribute to the perturbative
corrections to $Q_5$ (where now the dot refers to ${\cal O} = Q_5$). 
The Coulomb loop 
diagrams (Figs. 3.g and h) are zero because $Q_5$
does not contain terms proportional to $b^\dagger d^\dagger$ or $bd$.
Doing the requisite spin sums and performing the integral over loop momentum
and ${\bf k}$ leads to the same integrals for Figs. 3.a and 3.b as in 
Eqs. [52] and [53]. 
These are cancelled by the vertex correction integral of Fig. 3.c.  
The same holds for the Z-graphs of Figs. 3.d,e, and f.
Thus there is no correction to $Q_5$ at order $g^2$.
This is as
expected because the axial current is partially conserved and carries no
anomalous dimension.

\hbox to \hsize{%
\begin{minipage}[t]{\hsize}
\begin{figure}
\epsfxsize=5in
\hbox to \hsize{\hss\epsffile{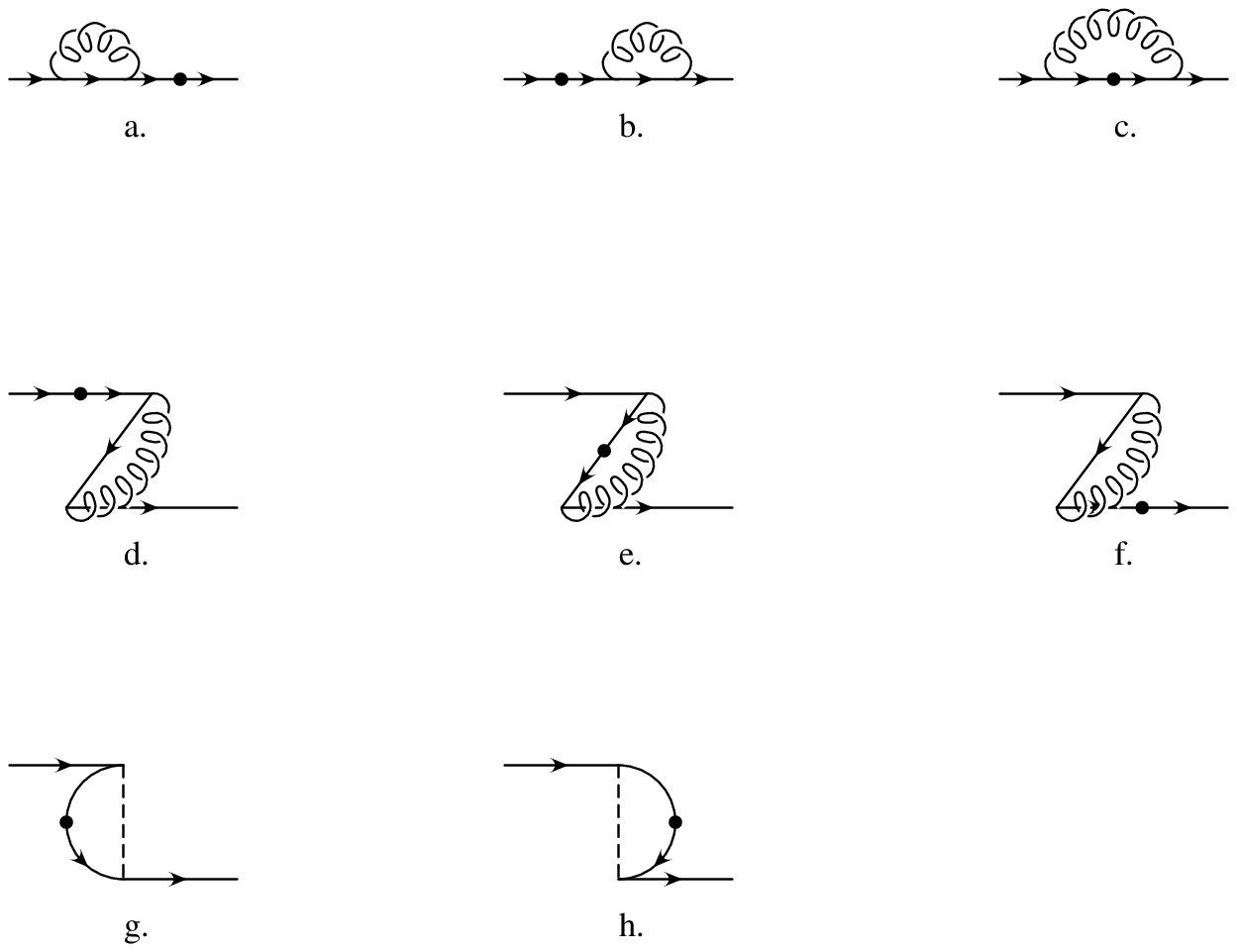}\hss}
\label{fig:Oeff}
\end{figure}
\end{minipage}}
\begin{center}
  {\small Fig.~3. Diagrams Contributing to ${\cal O}_{eff}$.}
\end{center}
\subsection{Renormalization of $\bar qq$}

The evaluation of Eq.~[\ref{eq:Oeff}] in the case of the $\bar q q$ operator is 
more involved than in the case of the axial charge because diagrams
which correspond to $b^\dagger b$ and $d^\dagger d$ also contribute.
The regulated operator $\bar q q^{\Lambda_0}$ is 

\begin{eqnarray}
& & \bar q q^{\Lambda_0} = \sum_\tau\int {{d\q} \over {(2\pi)^3}}
\phantom{-} \left[ s(\q) 
 [b^{\dag}(\q,\tau)b(\q,\tau) + d^{\dag}(-\q,\tau)d(-\q,\tau)] 
\right.
\nonumber \\
& & \phantom{ \bar q q^\Lambda_0 = \sum_\tau\int {{d\q} \over {(2\pi)^3}} }
\left. - c(\q)\theta({\Lambda_0 \over 2} - E(\q))
 [b^{\dag}(\q,\tau)d^{\dag}(-\q,\tau) + d^{\dag}(-\q,\tau)b(\q,\tau)] \right] 
\nonumber \\
& & \phantom{  \bar q q^{\Lambda_0} = } 
 - \sum_{\tau} \int { {d\q} \over {(2\pi)^3}} s(\q) 
\label{qq}
\end{eqnarray}
 where  $s(\q)$ and $c(\q)$ are given by

\begin{eqnarray}
s(\q) &=& {m_0 \over \sqrt{q^2 + m_0^2}} \\
c(\q) &=& {|\q| \over \sqrt{q^2 + m_0^2}} .
\end{eqnarray}

After similarity transformation 
the effective operator $\bar q q^{\Lambda_1}_{eff}$ can be written as 
\begin{eqnarray}
\bar q q^{\Lambda_0} &=&  \bar q q^{\Lambda_1} + g^2 {{C_F}\over {(4\pi)^2}}
\sum_\tau\int {{d\k} \over {(2\pi)^3}} \Big[ \phantom{-}
A_s(\k;\Lambda_0,\Lambda_1)
[b^{\dag}(\k,\tau)b(\k,\tau) + d^{\dag}(-\k,\tau)d(-\k,\tau)] 
\nonumber \\
&-& A_c(\k;\Lambda_0,\Lambda_1)
[ b^{\dag}(\k,\tau) d^{\dag}(-\k,\tau) + d(-\k,\tau) b(\k,\tau)] \Big]
\end{eqnarray}
Recall that $\bar q q^{\Lambda_0} = \bar q q^{\Lambda_1}_{eff}$ by definition.
The integrands in this expression may be separated into portions 
representing contributions from $[R_2,\bar q q]$ and 
$[R_1,[R_1,\bar q q]]$ respectively: 

\begin{equation}
A_s = A_{s2} + A_{s11},\;\; A_c = A_{c2} + A_{c11}.
\end{equation}
The commutator $[R_1,\bar q q^{\Lambda_0}]$ 
leads to an effective operator that changes the number of gluons by $\pm 1$ 
and hence need not be considered. 
The parts of these commutators which give rise to the leading renormalization group 
trajectory are given by 

\begin{eqnarray}
A_{s2}(\k;\Lambda_0,\Lambda_1) &=& \phantom{+} s(\k) 
 \log{{\Lambda_0}\over 2}
 \left[ 8 - 4\theta( |\k| + E(\k) - \Lambda_1)\left(1 - 
{{\Lambda_1 - E(\k)}\over {|\k|}}  \right) \right]  \nonumber \\
 &+& s(\k) \int d|\q|d(\hat\q\cdot\hat\k) \theta({\Lambda_0\over 2} - |\q|)
\Bigg[ {{\theta(\Lambda_0 > |-E(\k)-E(\q)+\omega(\k-\q)| > \Lambda_1)}
\over { -E(\k)-E(\q)+\omega(\k-\q) }} 
\nonumber \\
%&& 
%\phantom{ + s(\k) \int d|\q|d(\hat\q\cdot\hat\k) \theta({\Lambda_0\over 2}
%- |\q|) }
 &+ & {{\theta(\Lambda_0 > |E(\k)-E(\q)+\omega(\k-\q)| > \Lambda_1)}
\over { E(\k)-E(\q)+\omega(\k-\q) }} \Bigg].
 \end{eqnarray}
and
\begin{eqnarray}
A_{c2}& & (\k;\Lambda_0,\Lambda_1) = \phantom{+} c(\k) 
 \log{{\Lambda_0}\over 2}
 \left[ 8 - 4\theta( |\k| + E(\k) - \Lambda_1)\left(1 - 
{{\Lambda_1 - E(\k)}\over {|\k|}}  \right) \right]  \nonumber \\
&&+ \int  d|\q|d(\hat\q\cdot\hat\k)
\theta({\Lambda_0\over 2} - |\q|)
\Bigg[ 
(c(\k)+\hat\k\cdot\hat\q)
 {{\theta(\Lambda_0 > |-E(\k)-E(\q)+\omega(\k-\q)| > \Lambda_1)}
\over { -E(\k)-E(\q)+\omega(\k-\q) }} 
\nonumber \\
%&& 
%\phantom{+ d|\q|d(\hat\q\cdot\hat\k)
%\theta({\Lambda_0\over 2} - |\q|)}
 &&+ (c(\k)-\hat\k\cdot\hat\q)
 {{\theta(\Lambda_0 > |E(\k)-E(\q)+\omega(\k-\q)| > \Lambda_1)}
\over { E(\k)-E(\q)+\omega(\k-\q) }} \Bigg]. \nonumber \\
\end{eqnarray}

\noindent
The remaining terms are 
\begin{equation}
 A_{s11}(\k;\Lambda_0,\Lambda_1) = -2  s(\k) 
 \log{{\Lambda_0}\over 2} - s(\k) \int d|\q|d(\hat\q\cdot\hat\k) [...]
\end{equation}
and
\begin{equation}
 A_{c11}(\k;\Lambda_0,\Lambda_1) = -2  c(\k) 
 \log{{\Lambda_0}\over 2} -  \int d|\q|d(\hat\q\cdot\hat\k) [...].
\end{equation}
The two integrands in $A_{s11}$ and $A_{c11}$ abbreviated by $[...]$ are 
identical with the integrands in $A_{s2}$ and $A_{c2}$ respectively and 
therefore cancel in $A_s$ and $A_c$.
The contributions to $A_{s2}$ and $A_{c2}$ which are proportional to 
$\theta( |\k| + E(\k) - \Lambda_1)$ vanish when matrix elements of 
$\bar q q^{\Lambda}_{eff}$ are taken 
 in a basis in which  
$H^{\Lambda_1}_{eff}$ is  nonzero {\it i.e.}, for  
 $\Lambda_1 > 2E(\k)$ (see Eq.~[\ref{below}]). 
The additional contribution to $R_2$ resulting from off-diagonal self 
energies ($G_C$, $G_T$),  discussed below Eq.~[\ref{hct}],  
 results in  terms proportional to 
$\theta(2E(\k) - \Lambda_1)$ and also vanishes below the cut-off. 
Finally, after a single application of the similarity transformation, 

\begin{equation}
\bar q q^{\Lambda_0}
= \left[1 + g^2{{6C_F} \over 
 {(4\pi)^2}}\log{{\Lambda_0} \over {\Lambda_1}} \right] \bar q q{^\Lambda_1}  
 + \mbox{``finite''}
 =  \left[1 + g^2{{6C_F} \over 
 {(4\pi)^2}}\epsilon\left(1 + O\left({1\over {e^{N\epsilon}}}\right) \right)
 \right] \bar q q^{\Lambda_1}. \label{qqrenor}
\end{equation}
The term ``finite'' refers to contributions which are finite in the limit 
 $\Lambda_0 \to \infty$, with fixed $\Lambda_1$, and thus do not contribute 
 to the leading renormalization group trajectory. 
Applying the similarity transformation $N \sim \log {\Lambda_0/\Lambda_N}$ 
times  yields, 
\begin{equation}
\bar q q^{\Lambda_0} = 
\left( {\Lambda_0 \over \Lambda_N} \right)^{6C_F{{g^2}\over {(4\pi)^2}}}
 \bar q q^{\Lambda_N}. 
\end{equation}
Using Eq.~[\ref{massrun}] we may easily show that 
 the product 
$m \bar q q$, is constant along the renormalization 
group trajectory {\it i.e.}, as $\Lambda_0 \to \Lambda_N$ 

\begin{equation}
m(\Lambda_0)\bar q q^{\Lambda_0} = m(\Lambda_N)\bar q q^{\Lambda_N}.
\end{equation}
%The finite pices above depend on both the renormalization condition and 
% renormalization scale. In Fig. XX the effect of the similarity 
%trnasformation 

\section{Nonperturbative Properties}

It is known that constituent quarks are the appropriate degrees of freedom for
the description of hadrons. Since the DQM is constructed with current quarks, 
constituent quarks must be dynamically generated as the pseudoparticles
of the model. We therefore choose to proceed by making a pairing ansatz for
the vacuum of the standard BCS form. This is motivated by
the constituent quark model as we shall see shortly. The incorporation of a 
nontrivial pairing vacuum may be easily achieved by performing a 
Bogoliubov-Valatin canonical transformation on the quark fields:
\begin{eqnarray}
B(\k,\tau) &=& \cos(\h \phi(\k)) b(\k,\tau) - h(\tau) \sin(\h \phi(\k)) 
 d^{\dag}(\k,\tau) \nonumber \\
D(\k,\tau) &=& \cos(\h \phi(\k)) d(\k,\tau) + h(\tau)\sin(\h \phi(\k)) 
b^{\dag}(\k,\tau)
 \label{BV}
\end{eqnarray}
where $h(\tau)=\pm 1$ is the helicity. 
The transformation is parameterized in terms of the Bogoliubov angle, 
$\phi(\k)$. In future we shall use the notation

\begin{equation}
S_\h(\k) = \sin(\h\phi(\k)), \;\; S(\k) = \sin(\phi(\k)),
\end{equation}
with similar expressions for the cosine.

Expressing the effective Hamiltonian, $H^\Lambda_{eff}$  
in terms of the constituent 
quark and antiquark operators ($B$ and $D$ respectively) leads 
to a one-body off-diagonal operator of the type 
 
\begin{equation}
 \sum_{\tau} \int {{d\k} \over {(2\pi)^3}} F(\k;\phi)
\left[ B^{\dag}(\k,\tau)D^{\dag}(-\k,\tau) + D(-\k,\tau)B(\k,\tau) \right].
\end{equation}
The Bogoliubov angle is chosen so that this operator vanishes
\begin{equation}
F(\k,\phi) = 0; \label{gap0}
\end{equation}
this is the vacuum gap equation of the DQM.

Although the gap equation allows the determination of a nontrivial 
vacuum, and hence of the dynamical quark mass (discussed below), there is 
actually another important reason for imposing this structure on the model.
In particular,
the successes of the constituent quark model indicate that valence
quarks saturate hadronic states to a large degree. The 
BCS ansatz is one of the simplest methods which allows for dynamical 
chiral symmetry breaking and decoupling of $\bar q q$ pairs from the 
 vacuum. 
 Thus the nonperturbative approach
we have taken is firmly based on the successes of the CQM. We note  that
the decoupling of the vacuum does not persist as states with four or 
more constituents are included in the analysis. In this case
it is only possible 
to approximately diagonalize the Hamiltonian 
by further changing the basis as it is done, for example in the 
Random Phase Approximation (RPA) or by 
using bound state perturbation theory with either Tamm-Dancoff or 
RPA eigenstates as a basis. 

The explicit form of the gap equation is given by:

\begin{eqnarray}
 & &  \left[ \sqrt{\k^2+m(\Lambda)^2} + \Sigma_l(\k,\Lambda) \right]  
S(\k)  + \theta({\Lambda\over 2} - E(\k)) 
 G_l(\k;\Lambda) 
 C(\k)  
 = \nonumber \\ 
& & =\int {d\q \over (2 \pi)^3}\left[  \left( {\cal V}^{(1)}_{cl}(\k,\q) + {\cal V}^{(1)}_T(\k,\q) \right) 
  2 S(\k) S_\h^2(\q) 
+   \left( {\cal V}^{(2)}_{cl}(\k,\q) + {\cal V}^{(2)}_{T}(\k,\q) \right) 
C(\k) S(\q) \right.\nonumber \\
& & +\left. \left( {\cal V}^{(3)}_{cl}(\k,\q) + {\cal V}^{(3)}_{T}(\k,\q) \right) 
S(\k) S(\q)
+ \left( {\cal V}^{(4)}_{cl}(\k,\q) + {\cal V}^{(4)}_{T}(\k,\q) \right) 
2 C(\k) S_\h^2(\q)\right]. \label{gap}
\end{eqnarray}          
Here  the functions ${\cal V}^{(i)}_{cl}$ and ${\cal V}^{(i)}_{T}$ come from 
the static, two-body,  Coulomb and linear (denoted by the subscript $cl$)
part of the effective Hamiltonian and the effective interactions resulting 
from transverse gluon exchange (denoted by $T$)
respectively. They are all summarized in the Appendix. 
The only self-energy type contributions, $\Sigma_l$ and $G_l$, come 
from the linear potential. As discussed in 
Sec.~IID, these are necessary to 
maintain the IR stability of physical observables. The importance of the
linear self energy term has been noted previously in Refs. \cite{ssjc,AD,port}.
It should be noted that the solution to the gap equation
depends on the scale $\Lambda$, $\phi = \phi(\k;\Lambda)$. 

Since the gap equation incorporates all order $g^2$ physics correctly it is
completely UV finite. This is not true for earlier work. For example, 
the gap equation of Le Yaouanc {\it et al.}
\cite{Orsay} may by obtained from this one by setting all theta functions 
equal to unity and by ignoring 
all transverse gluon terms.  This introduces an ultraviolet divergence due
to the Coulomb self-energy term which they avoid by simply neglecting the 
Coulomb interaction. 
 Adler and Davis \cite{AD} discuss the gap equation with a 
pure Coulomb interaction. It should be noted, however, that in absence of 
transverse gluons, 
pure Coulomb exchange leads to momentum dependent counter-terms.
Finally, the gap equation
of Finger and Mandula \cite{FM} is  similar to those above except that they
ignore all self energy terms. This eliminates all UV divergences but introduces
IR divergences.  We stress that the only consistent way to derive a gap
equation which incorporates gluons is with a well-defined Hamiltonian-based
renormalization scheme, such as is presented here.

As $|\k| \to \infty$ the solution to the gap equation approaches the 
perturbative result: $S(\k) \to s(\k)$. It is thus natural to define
an effective quark mass with the relationship

\begin{equation}
s(\k;m_{dyn}(\k)) = S(\k).
\end{equation}

\noindent
The dispersion relation for such a dynamical quark mass is shown in Fig. 4 
(the bare quark mass was chosen to satisfy $m$(1 GeV)$ =$ 5 MeV).
The solid line is $E(k;\Lambda)$ for $\Lambda = 4$ GeV. The dashed line
represents $\sqrt{k^2 + m^2(4 {\rm GeV})}$ (where the quark mass
has been perturbatively run from 1 GeV to 4 GeV).
One
sees that the correct high energy behavior is recovered and that a constituent
quark mass of roughly 100 MeV is obtained at low energy. This is approximately
one half of the constituent masses used in relativistic quark models. 
In general the constituent quark mass will tend to zero as the scale is 
reduced and will saturate as the scale is increased. A detailed study of the
dynamical quark mass is beyond the scope of this paper and will be presented 
elsewhere.

\hbox to \hsize{%
\begin{minipage}[t]{\hsize}
\begin{figure}
\epsfxsize=5in
\hbox to \hsize{\hss\epsffile{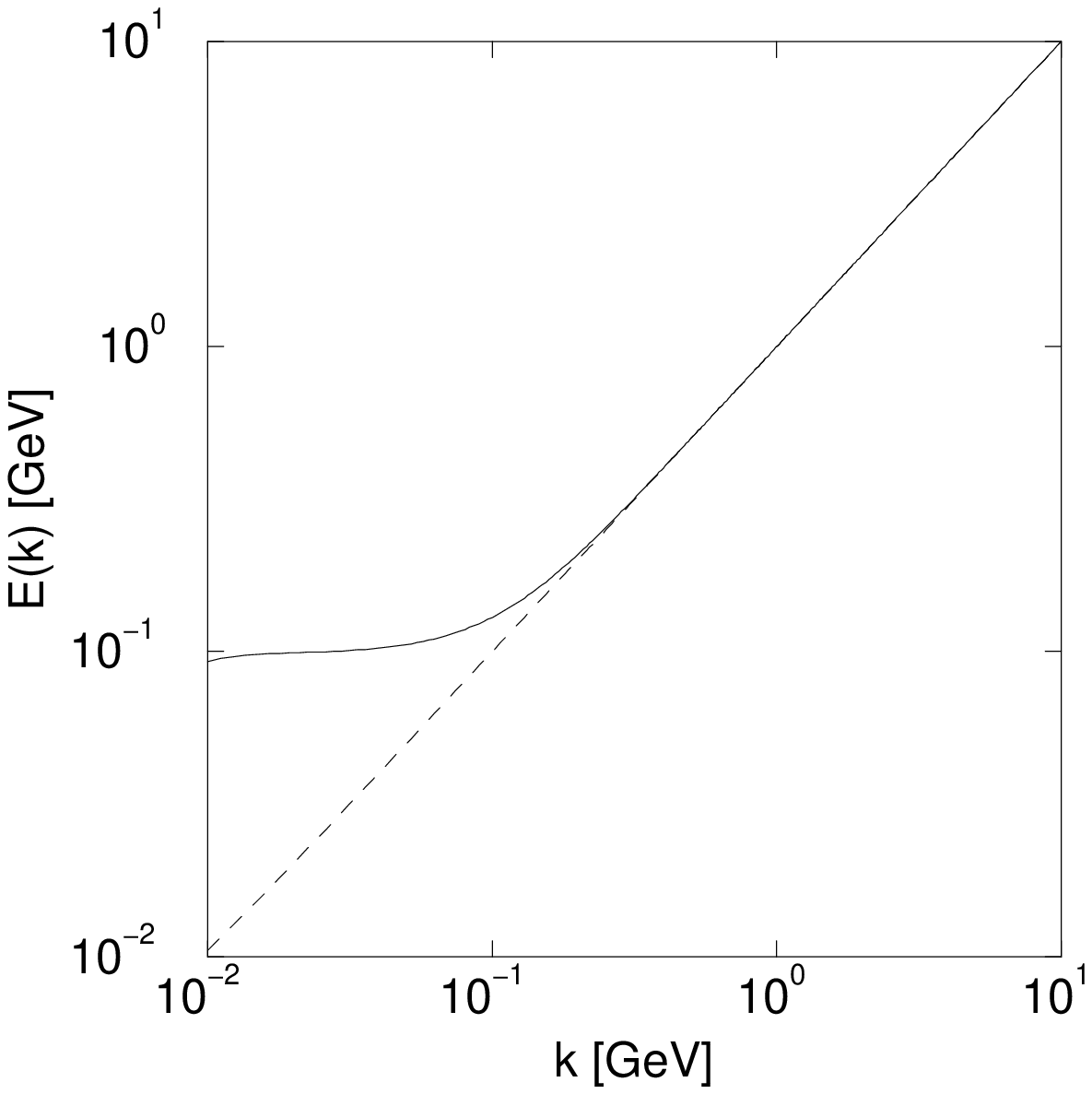}\hss}
\label{fig:mdyn}
\end{figure}
\end{minipage}}
\begin{center}
  {\small Fig.~4. The Dynamical Quark Dispersion Relation}
\end{center}

With a nontrivial solution to the gap equation we may proceed to define the 
nonperturbative quark condensate. 
The  matrix element of $\bar q q^\Lambda_{eff}$ in the 
nonperturbative vacuum, $|\Omega \rangle$ (defined by 
$B|\Omega \rangle = D|\Omega \rangle = 0$) 
is given by (per quark flavor)

\begin{equation}
\langle \Omega| \bar q q^\Lambda | \Omega \rangle = - 6
 \int {{d\k} \over {(2\pi)^3}} \left[ S(\k) c(\k) + (C(\k) - 1) s(\k) \right].
\end{equation}

\noindent
This expression has been regulated by subtracting the perturbative contribution:

\begin{equation}
\langle 0 | {\bar q} q | 0 \rangle = -6 \int {d\k \over (2 \pi)^3} s(k).
\end{equation}

\noindent
The condensate is shown as a function of the renormalization scale in Fig. 5.
The circles represent the solution for massless quarks while the diamonds
are for $m(1 {\rm GeV})= 5$ MeV. The perturbative solution is shown as a solid
line (this has been normalized to match the nonperturbative results). As 
expected, both nonperturbative curves approach the perturbative solution as
$\Lambda$ increases. The behavior of the condensate below 3 GeV is driven by 
the nonperturbative renormalization group trajectory of the model. How 
closely this
follows the actual nonperturbative evolution of QCD is a function of how good
our approximations are (in particular the form of $V_{conf}$ and the vacuum 
Ansatz). A way to test the consistency of the model is to calculate the nonperturbative running quark
mass and strong coupling constant (and wave function normalization). This may be
done, {\it e.g.}, by fixing $\Lambda$ (below 3 GeV), calculating several 
observable (such
as the pion and rho masses) and fixing the parameters of the model to reproduce 
experiment,

\begin{equation}
m_{\pi/\rho}(m(\Lambda),\alpha_S(\Lambda);\Lambda) = m_{\pi/\rho}|_{\rm expt}.
\end{equation}

\noindent
Once the nonperturbative running mass is determined one may compare it to 
the expected running mass (taking advantage of the renormalization group
invariance of $m \, q \bar q$ and the results of Fig. 5.) to
assess the accuracy of the model.

The value of the condensate at low energy is roughly ($-$100 MeV)$^3$. This does
not compare well with the commonly quoted value of ($-$250 MeV)$^3$ \cite{SVZ}.
The underestimate for both the constituent quark mass and the condensate 
is most likely
due to an inadequate model for the vacuum. More quark correlations may need
to be incorporated or a coupled quark-gluon vacuum Ansatz may be required.
Unfortunately the use of nonperturbative methods implies that the problem
is not restricted to the vacuum sector. For example, the Thouless theorem
relates the pairing vacuum to the RPA pion via the 
Gell-Mann--Oakes--Renner relationship $f_\pi^2 m_\pi^2 = -2 m_q 
\langle q \bar q \rangle$\cite{ssjc}. Thus, if the pion mass is fixed, a poor 
model of the vacuum will be reflected in a low value of the pion decay constant. How severe these problems are will be examined in future work. For now, we
must remain satisfied that nonperturbative statements about the vacuum can
be made at all.

\hbox to \hsize{%
\begin{minipage}[t]{\hsize}
\begin{figure}
\epsfxsize=5in
\hbox to \hsize{\hss\epsffile{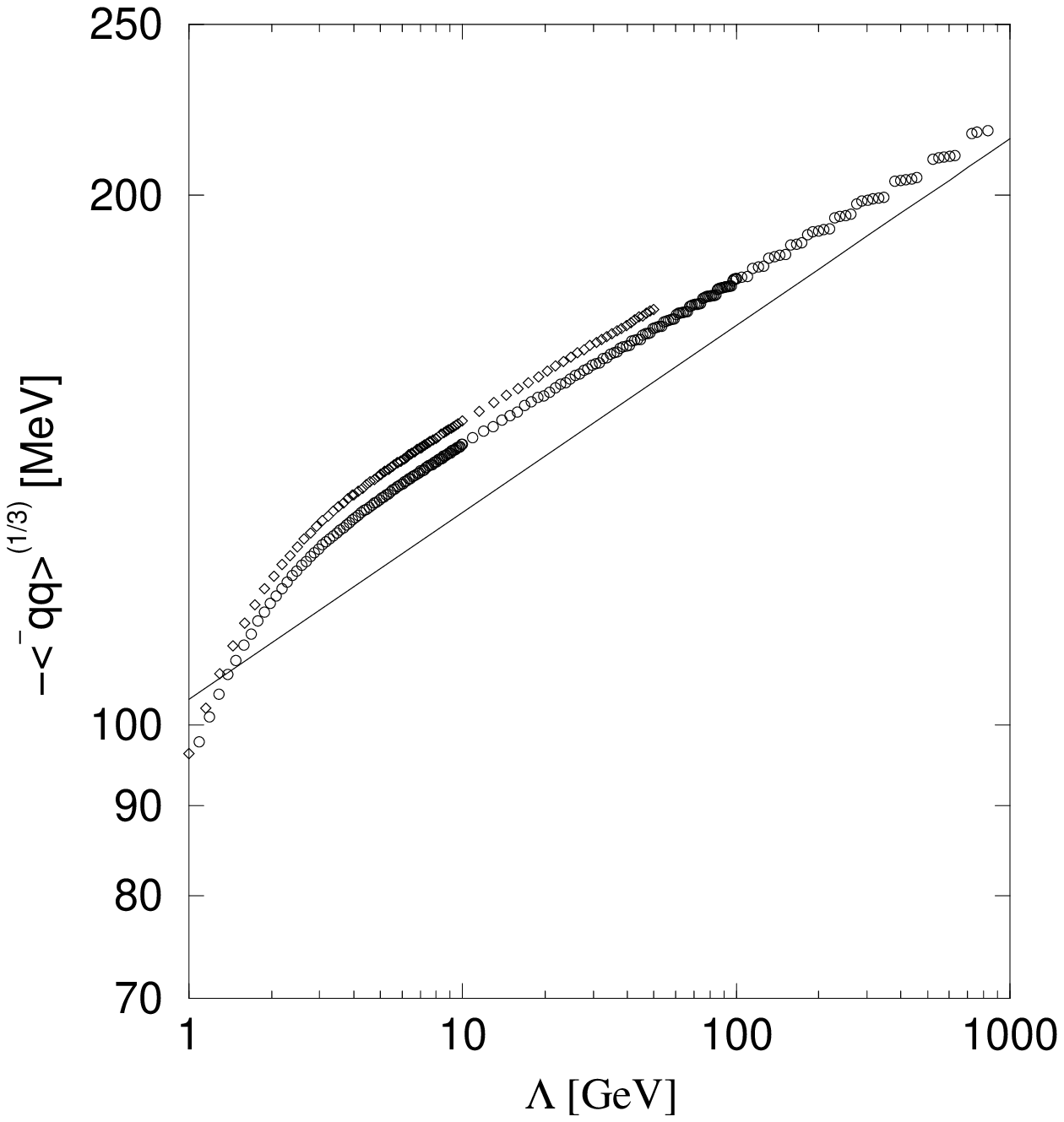}\hss}
\label{fig:cond}
\end{figure}
\end{minipage}}
\begin{center}
  {\small Fig.~5. The Quark Condensate.}
\end{center}

\section{Conclusions}

This paper presents the development of the Dynamical Quark Model. The model
is meant to describe hadronic physics and is therefore designed to be
amenable to many-body techniques from the outset. The use of field theoretic 
methods 
forces one to choose appropriate degrees of freedom and many-body 
approximations. The 
constituent quark model has been used as a guide in this process. For example,
the choice of the Coulomb gauge, the form of the gap equation, and the use
of the Tamm-Dancoff approximation are all suggested
by the phenomenological success of the CQM.

There are three main stages in the construction of the DQM. The first is 
the evaluation of the renormalization group improved QCD Hamiltonian to order
$g^2$. This is achieved with a Hamiltonian-based renormalization scheme 
introduced by G{\l}azek and Wilson (the similarity scheme). 
In the second stage, the RGI Hamiltonian is supplemented with a 
phenomenological potential. The fact that the starting point of the DQM is
QCD provided some important constraints. In particular, the Dirac structure of
the phenomenological potential is determined in the heavy quark limit. 
Performing the Bogoliubov-Valatin transformation comprised the last stage of
this program. This allows the construction of a nontrivial vacuum -- a necessary
feature of the model since constituent quarks must be dynamically generated. A
beneficial corollary is the appearance of chiral pions. 

Since this paper is meant to describe the perturbative development of the 
model, only the simplest preliminary applications have been discussed.
Thus we have derived and solved the vacuum gap equation, used the resulting
Bogoliubov angle to determine the dynamical quark mass, and evaluated the
chiral quark condensate. It is satisfying and nontrivial that a constituent
quark mass arises and is roughly as expected from phenomenological 
considerations. This is 
an indication that the approach we are taking may be useful for light quarks
(the DQM may be trusted for heavy quarks since it reduces
to the CQM in this case). Unfortunately, the chiral condensate and the
constituent quark mass are too small --
evidence that the nonperturbative structure of QCD is not fully captured in
our model and approximations. Whether this is a serious problem or not can
only be determined by a comparison with more observables. The application
(and possible modification) of the DQM to hadronic observables is clearly an
important task for the future.

The DQM has a rich structure and therefore allows the examination of many 
topical 
issues in hadronic physics. There are several qualitative aspects of the model
which we hope to explore in the near future. These include an extension
of the phenomenological potential to incorporate nonperturbative gluonic 
flux tube
degrees of freedom and an analysis of hadronic decays in a spirit similar to
our discussion
of the dynamical generation of effective scalar spin-dependent 
interactions\cite{ss2}. It will also be interesting to examine hyperfine
splittings in the N-$\Delta$ and $\pi$-$\rho$ systems as a function of quark mass.
This should shed some light on the genesis and utility of the constituent
quark model.

Finally, there is a wealth of specific phenomenological problems in hadronic
physics which the DQM may address. For example, since gluonic degrees of 
freedom are explicitly included in the model, we may examine the 
properties of glueballs and hybrids -- including their couplings to quarkonia.
Pions are ubiquitous in hadronic physics, appearing as strongly interacting  
probes, decay products, and exchange currents. It is therefore essential that 
they be thoroughly understood. However the chiral and relativistic nature of
pions have made this a longstanding deficiency of microscopic models of QCD.
It will therefore be interesting to examine a selection of problems which 
involve pions (recall that pions are relativistic pseudo-Goldstone
bosons built from quark pseudoparticles in the DQM). In particular we plan
to calculate the pion mass, decay constant, electromagnetic form factor, and
the width for $\pi_0 \to \gamma\gamma$. Many other problems may be addressed
with the DQM. We regard the scheme presented here as the next step in a 
process where models are used to elucidate and guide experiment which in
turn directs the development of newer and more sophisticated models of low 
energy QCD.

\acknowledgements
Financial support from the U.S. Department of Energy grants No. 
DE-FG05-88ER40461, DE-FG05-90ER40589, DE-FG02-96ER40944 and Cray time from 
the North Carolina Supercomputer Center is acknowledged.   
ES also acknowledges the financial support of an  
FRPD grant from NCSU.

\appendix
\section{}

The functions $\Sigma_C$ and $G_C$ appearing in Eq.~[\ref{vc}] are given by 

\begin{equation}
\Sigma_C(\k;\Lambda^C_0) = g^2 C_F  \int {{d\q}\over {(2\pi)^3}}
\theta(\Lambda^C_0 - E(\q))
\left( s(\k)s(\q) + \hat\k\cdot\hat\q c(\k)c(\q)\right)V_c(\k,\q)
\end{equation}

\begin{equation}
G_C(\k;\Lambda^C_0) = -  g^2 C_F  \int {{d\q}\over {(2\pi)^3}}
\theta(\Lambda^C_0 - E(\q))
\left( c(\k)s(\q) - \hat\k\cdot\hat\q s(\k)c(\q)\right)V_c(\k,\q)
\end{equation}
where

\begin{equation}
V_c(\k,\q) = {1\over {2|\k - \q|^2}}.
\end{equation}

The self energies due to transverse gluon exchange are given by

\begin{eqnarray}
& & \Sigma_T(\k;\Lambda_0,\Lambda_1) = 
-g^2 C_F \int {{d\q}\over {(2\pi)^3}}
\left(1 - s(\k)s(\q) - \hat\k\cdot\hat\bl\hat\q\cdot\hat\bl c(\k)c(\q)
\right) \nonumber \\
& & \times { { \theta(\Lambda_0>|E(\k)-E(\q)-\omega(\bl)|>\Lambda_1) }\over 
{2\omega(\bl)\left(\omega(\bl)+E(\q)-E(\k)\right)}} 
 + 
\left(-1 - s(\k)s(\q) - \hat\k\cdot\hat\bl\hat\q\cdot\hat\bl c(\k)c(\q)
\right) \nonumber \\ 
& & \times { { \theta(\Lambda_0>|E(\k)+E(\q)+\omega(\bl)|>\Lambda_1) }\over 
{2\omega(\bl)\left(\omega(\bl)+E(\q)+E(\k)\right)}} \nonumber \\
\end{eqnarray}

\noindent
and 

\begin{eqnarray}
& & G_T(\k;\Lambda_0,\Lambda_1) =  g^2 C_F  \int {{d\q}\over {(2\pi)^3}}
{1 \over {\omega(\bl)}}
\theta(\Lambda_0 - |E(\q)+E(\k)+\omega(\bl)|)
\theta(\Lambda_0 - |E(\q)-E(\k)+\omega(\bl)|) \nonumber \\ 
& & \times 
\left( c(\k)s(\q) - \hat\k\cdot\hat\bl\hat\q\cdot\hat\bl c(\q)s(\k)\right)
\Theta_T(E(\q)+E(\k)+\omega(\bl),E(\q)-E(\k)+\omega(\bl);\Lambda_1).
\nonumber \\
\end{eqnarray}

The self energies due to the linear potential appearing in Eq.~[\ref{gap}], 
$\Sigma_l$ and $G_l$, are obtained from $\Sigma_c$ and $G_c$ respectively 
by substitution 

\begin{equation}
 g^2 C_F  V_c(\k,\q) \to V_l(\k,\q) = {{4\pi \sigma }\over {|\k-\q|^4}},
\end{equation}
where $\sigma = 0.18\mbox{ GeV}^2$ is the string tension.

The functions  ${\cal V}^{(i)}_{cl}$ which appear in  Eq.~[\ref{gap}]
are given by

\begin{eqnarray}
& & {\cal V}^{(1)}_{cl}(\k,\q) = \left(s(\k) s(\q) + \hat\k\cdot\hat\q c(\k) s(\q)\right) 
V_{cl}(\k,\q) \nonumber \\
& & {\cal V}^{(2)}_{cl}(\k,\q) = \phantom{+}{1\over 2}\left( c(\k) c(\q) 
+ \hat\k\cdot\hat\q(1+s(\k) s(\q)) \right)
\Theta_-(\k,\q;
\Lambda)V_{cl}(\k,\q) \nonumber \\
& &  \phantom{ {\cal V}^{(2)}_{cl}(\k,\q) = } + 
{1\over 2}\left( c(\k) c(\q) - \hat\k\cdot\hat\q(1-s(\k) s(\q)) \right)\Theta_+(\k,\q;
\Lambda)V_{cl}(\k,\q) \nonumber \\
& & {\cal V}^{(3)}_{cl}(\k,\q) = \left(-s(\k) c(\q) + \hat\k\cdot\hat\q s(\q) c(\k) 
\right) \Theta_\q(\q;\Lambda)V_{cl}(\k,\q)
\nonumber \\
& & {\cal V}^{(4)}_{cl}(\k,\q) = {\cal V}^{(3)}_{cl}(\q,\k) 
\end{eqnarray}

\noindent
and

\begin{eqnarray}
& & {\cal V}^{(1)}_T(\k,\q) = \phantom{+}  {1\over {4\omega(\bl)}}
\left[ 1 - s(\k) s(\q)
 - \hat\k\cdot\hat\bl \hat\q\cdot\hat\bl c(\k) c(\q) \right] \nonumber \\
& & 
\phantom{  {\cal V}^{(1)}_T(\k,\q) =  + }
\times 
[\phantom{+}\Theta_T(E(\q)-E(\k)+\omega(\bl),
 E(\q)-E(\k)+\omega(\bl);\Lambda) 
  \nonumber \\
& &  \phantom{ {\cal V}^{(1)}_T(\k,\q) =  +
 \times 
[  }
 + \Theta_T(-E(\q)+E(\k)+\omega(\bl),
 -E(\q)+E(\k)+\omega(\bl)
;\Lambda) ] \nonumber \\
&  & \phantom{  {\cal V}^{(1)}_T(\k,\q) = }
-{1\over {4\omega(\bl)}} \left[ 1 + s(\k) s(\q) + \hat\k\cdot\hat\bl\hat\q
\cdot\hat\bl c(\k) 
c(\q) \right] \nonumber \\
& & \phantom{  {\cal V}^{(1)}_T(\k,\q) = + }
\times [\phantom{+}\Theta_T(-E(\q)-E(\k)+\omega(\bl),
-E(\q)-E(\k)+\omega(\bl);\Lambda) \nonumber \\
 & &  \phantom{  {\cal V}^{(1)}_T(\k,\q) = +  \times [ }
+ \Theta_T(E(\q)+E(\k)+\omega(\bl),
 E(\q)+E(\k)+\omega(\bl)
;\Lambda) ]
\nonumber \\
\end{eqnarray}
\begin{eqnarray}
& & {\cal V}^{(2)}_T(\k,\q) = - {1\over {2\omega(\bl)}}\left[ c(\k) c(\q) 
- \hat\k\cdot\hat\bl\hat\q\cdot\hat\bl(1-s(\k) s(\q)) \right]
\Theta_-(\k,\q;\Lambda)\nonumber \\
& & \phantom{ {\cal V}^{(2)}_T(\k,\q) = -  {1\over {2\omega(\bl)}}  }
 \times \Theta_T(-E(\q)+E(\k)+\omega(\bl),E(\q)-E(\k)+\omega(\bl);\Lambda) 
\nonumber \\
& & \phantom{  {\cal V}^{(2)}_T(\k,\q) ={1\over {2\omega(\bl)}} }
- \left[ c(\k) c(\q) + \hat\k\cdot\hat\bl\hat\q\cdot\hat\bl(1+s(\k) s(\q))
 \right]
\Theta_+(\k,\q;\Lambda)\nonumber \\
& & \phantom{ {\cal V}^{(2)}_T(\k,\q) = -{1\over {2\omega(\bl)}} }
 \times \Theta_T(E(\q)+E(\k)+\omega(\bl),-E(\q)-E(\k)+\omega(\bl);\Lambda)
\nonumber \\
& & {\cal V}^{(3)}_T(\k,\q) = {1\over {2\omega(\bl)}}
\left[ s(\k) c(\q) - \hat\k\cdot\hat\bl\hat\q\cdot\hat\bl c(\k) s(\q)) \right]
\Theta_\q(\q,\Lambda) \nonumber \\
& & \phantom{  {\cal V}^{(3)}_T(\k,\q)   } 
\times [ \phantom{+}\Theta_T(-E(\q)-E(\k)+\omega(\bl),E(\q)-E(\k)+\omega(\bl)
;\Lambda) \nonumber \\
& & \phantom{  {\cal V}^{(3)}_T(\k,\q) \times (  } 
 +  \Theta_T(E(\q)+E(\k)+\omega(\bl),-E(\q)+E(\k)+\omega(\bl);\Lambda) ]
\nonumber \\
& & {\cal V}^{(4)}_T(\k,\q) =  {\cal V}^{(3)}_T(\q,\k)
\end{eqnarray}
where $V_{cl}$ is the sum of Coulomb and linear potentials 

\begin{equation}
V_{cl}(\k,\q) = {{C_F g^2} \over {2 |\k - \q|^2}} + {{4\pi \sigma}\over 
{|\k-\q|^4}}
\end{equation}

\noindent
and the effective 
transverse gluon cut-off $\Theta_T$ is  given by Eq.~[\ref{theta_T}].  
Additional cut-offs are given by 

\begin{equation}
\Theta_\pm(\k,\q;\Lambda) = \theta(\Lambda - 2|E(\k) \pm E(\q)|)
\end{equation}
and
\begin{equation}
\Theta_\p = \theta(\Lambda - 2E(\p)).
\end{equation}
Furthermore $s(\k)$ and $c(\k)$ stand for the sine and cosine 
 of the free particle BV angle {\it i.e.}, 

\begin{equation}
s(\p) = \sin(\phi_m(\p)) = {m\over \sqrt{\p^2 + m^2}},\;\; c(\p)= \cos(\phi_m(\p))= {1\over \sqrt{\p^2 + m^2}}.
\end{equation}
Finally, 
$\bl=\k-\q$ and $\omega(\bl) = |\bl|$
 represent the momentum and energy of an exchanged gluon respectively.

\newpage
\figure{ Fig.~1. 
 Resolving Ambiguities in Current Matrix Elements.}

\figure{Fig.~2. Diagrams Contributing to $b_{eff}$.}

\figure{Fig~.3. Diagrams Contributing to ${\cal O}_{eff}$.}

\figure{Fig.~4. 
 The Dynamical Quark Dispersion Relation. The solid line 
is the numerical result while the dashed line is the perturbative 
relation for $m$( 4 GeV) = 3.1 MeV.}

\figure{Fig.~5. 
 The Chiral Condensate. The solid line is perturbation theory, the
symbols represent the nonperturbative condensate for massless quarks (circles)
and for m(1 GeV) = 5 MeV (diamonds).}

\end{document}